\documentclass[aps,pre,a4paper,10pt,twocolumn,superscriptaddress,floatfix,longbibliography,nofootinbib]{revtex4-2}

\usepackage[english]{babel}
\usepackage[utf8]{inputenc}
\usepackage[T1]{fontenc}
\usepackage[nice]{nicefrac}
\usepackage{graphicx}
\usepackage[dvipsnames]{xcolor}
\usepackage{amsmath,amssymb,bm,bbm}
\usepackage{braket}
\usepackage[normalem]{ulem} 
\usepackage{orcidlink}
\usepackage{physics}

\definecolor{darkgreen}{rgb}{0,0.6,0.2}
\definecolor{darkblue}{rgb}{0.1,0.3,1}

\newcommand{\ie}{{\it i.e.},\ }

\usepackage{hyperref} 

\hypersetup{colorlinks, linkcolor=BrickRed, citecolor=MidnightBlue, urlcolor=blue}

\begin{document}
\title{Information in Many-body Eigenstates: A Question of Learnability}

\author{Maksymilian Kliczkowski\orcidlink{0000-0002-7987-9913}}
\affiliation{Center for Advanced Systems Understanding, Helmholtz-Zentrum Dresden-Rossendorf, Germany\looseness=-3}
\affiliation{Institute of Theoretical Physics, Faculty of Fundamental Problems of Technology, Wrocław University of Science and Technology, 50-370 Wrocław, Poland\looseness=-3}

\author{Jaros\l{}aw Paw\l{}owski\orcidlink{0000-0003-3638-3966}}
\affiliation{Institute of Theoretical Physics, Faculty of Fundamental Problems of Technology, Wroc\l{}aw University of Science and Technology, 50-370 Wroc\l{}aw, Poland\looseness=-3}

\author{Masudul Haque\orcidlink{0000-0001-6848-6068}}
\affiliation{Institut für Theoretische Physik, Technische Universität Dresden, 01062 Dresden, Germany\looseness=-3}


\begin{abstract}
To what extent do individual eigenstates encode information about their parent Hamiltonian, and how does this encoding vary across the spectrum?  We introduce \emph{learnability} as a new framework to quantify this information, measured by the precision with which a machine learning model can reconstruct a Hamiltonian from a limited set of eigenstates.  For many-body quantum systems, there is a contrast between the eigenstates near the spectral edges (low-entanglement, highly-structured states) and those far from the spectral edges (high-entanglement, near-random states). Using an encoder-decoder neural network for a non-integrable spin chain, we show that this dichotomy results in a stark difference in learnability: spectral-edge eigenstates allow for higher-accuracy Hamiltonian reconstruction using significantly fewer eigenstates, compared to mid-spectrum eigenstates.  Our results provide a new lens through which to view the spectral structure of many-body systems. 
 
\end{abstract}

\maketitle

\section{Introduction}\label{sec:intro}

Recent years have established machine learning (ML) as a promising framework for quantum many-body physics, with applications ranging from variational representation of quantum states~\cite{Carleo2017,lange_architecturesapplications_2024,medvidovic_neuralnetworkquantum_2024,dash_efficiencyneuralquantum_2025a,paul_boundentanglement_2026} or quantum state tomography~\cite{Torlai2018,Kotuny2022,balaz2025,krawczyk2026tomo}, to phase classification~\cite{broecker_machinelearning_2017,carrasquilla_machinelearning_2017,dong_machinelearning_2019,rem_identifyingquantum_2019,canabarro_unveilingphase_2019,shiina_machinelearningstudies_2020,yu_experimentalunsupervised_2021,mahlow_modelindependentquantum_2023,franco_quantumphases_2025} and material discovery~\cite{behler_generalizedneuralnetwork_2007,behler_perspectivemachine_2016,faber_machinelearning_2016,artrith_efficientaccurate_2017,seko_representationcompounds_2017,akinpelu_discoverynovel_2024,nematov_machinelearningdriven_2025}. It has also been used effectively to improve numerically demanding classical tasks~\cite{yoon_analyticcontinuation_2018,zhang_trainingbiases_2022,kliczkowski_autoencoderbasedanalytic_2024}.

Beyond practical applications, a promise of an emerging field is to provide new conceptualizations of physical phenomena. As an example from recent history, we recall how the concept of entanglement entropy entered condensed matter physics at the turn of century~\cite{Vidal_Latorre_Rico_Kitaev_PRL2003, Calabrese_Cardy_JSTAT2004, Korepin_PRL2004_entanglementscaling, Kitaev_Preskill_PRL2006, Levin_Wen_PRL2006}, opening up previously non-existent ways of thinking about many-body quantum physics. This included novel characterizations of phases, phase transitions, and dynamics~\cite{eisert2010colloquium, Amico_Fazio_Osterloh_Vedral_RMP2010_entanglement, Laflorencie_PhysRep2016_entanglement}. In this work, we address a phenomenon of many-particle physics by introducing a concept that only makes sense due to the accessibility of ML techniques for learning from the available data.

\begin{figure}[b]
    \centering
    \includegraphics[width=1.0\linewidth]{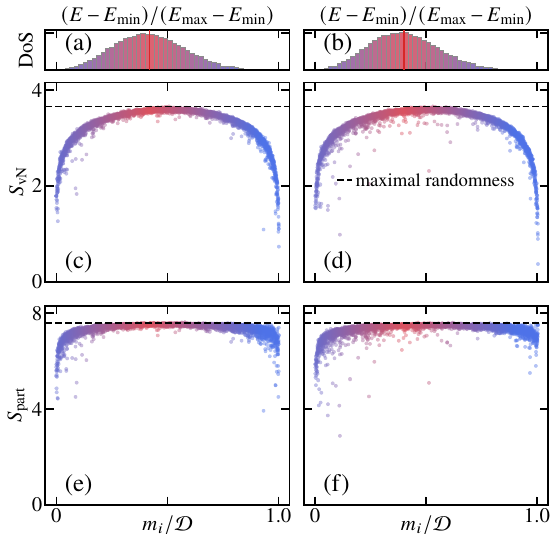}
    \caption{Typical distinction between eigenstates in different parts of the spectrum, demonstrated for the Hamiltonian \eqref{eq:hamil}, with $L=12$ sites, $\Delta=0.2$ and $J_1=-0.4$ ($J_1=0.4$) for the left (right) column.  (a),(b) Density of states (DoS), $\rho (E) = \sum _{m_i=1}^\mathcal{D} \delta(E-E_{m_i})$, as a function of rescaled energy.  (c),(d) bipartite (half-chain) entanglement entropy~\cite{eisert2010colloquium,bianchi_volumelawentanglement_2022a} $S_{\rm vN}$ [Eq.~\eqref{eq:vnentropy}], and (e),(f) participation entropy~\cite{luitz_participationspectroscopy_2014,vermersch_manybodyentropies_2024} $S_{\rm part}$ [Eq.~\eqref{eq:part:ent}]  versus normalized eigenstate index $m_i/\mathcal{D}$. A strong suppression is visible at the spectral edges.
    }
    \label{fig:fig0}
\end{figure}

The phenomenon we address is a difference between many-body eigenstates near the ground state and those far from it (we consider bounded Hilbert spaces, so we may select the extreme case of mid-spectrum eigenstates). It is widely appreciated that low-energy eigenstates of physical many-body Hamiltonians exhibit strong structure, described by locality~\cite{lieb_finitegroup_1972} or constrained entanglement~\cite{hastings_arealaw_2007,eisert2010colloquium}. On the other hand, mid-spectrum eigenstates are generally pseudo-random, consistent with random matrix theory (RMT) descriptions and the eigenstate thermalization hypothesis (ETH) built on top of RMT ideas~\cite{srednicki_94,deutsch_91,rigol_dunjko_08,dAlessio16,garrison_doessingle_2018a} with some system-dependent physical constraints~\cite{haque_entanglementmidspectrum_2022, huang_deviationmaximal_2024,kliczkowski_averageentanglement_2023a,langlett_entanglementpatterns_2025,patil_averagepurestate_2023,rodriguez_nieva_quantifyingquantum_2024,swietek_eigenstateentanglement_2024,yauk_typicalentanglement_2024}. 
In Figure \ref{fig:fig0}, we illustrate how the properties of eigenstates vary with energy by displaying the entanglement entropy [(c),(d)] and the participation entropy [(e),(f)] -- definitions are given in Appendix~\ref{app:entropy} for completeness. In the mid-spectrum region, where the density of states is large, the eigenstates are close to random states, and both properties shown ($S_{\text{vN}}$ and $S_{\text{part}}$) have values approaching the average random state value (dashed line).  The eigenstates at the spectral edges are non-generic and the properties are very different from random-state values.
This is a typical scenario -- the nature of eigenstates is highly sensitive to their spectral position.

\begin{figure*}[!ht]
    \centering
    \includegraphics[width=\textwidth]{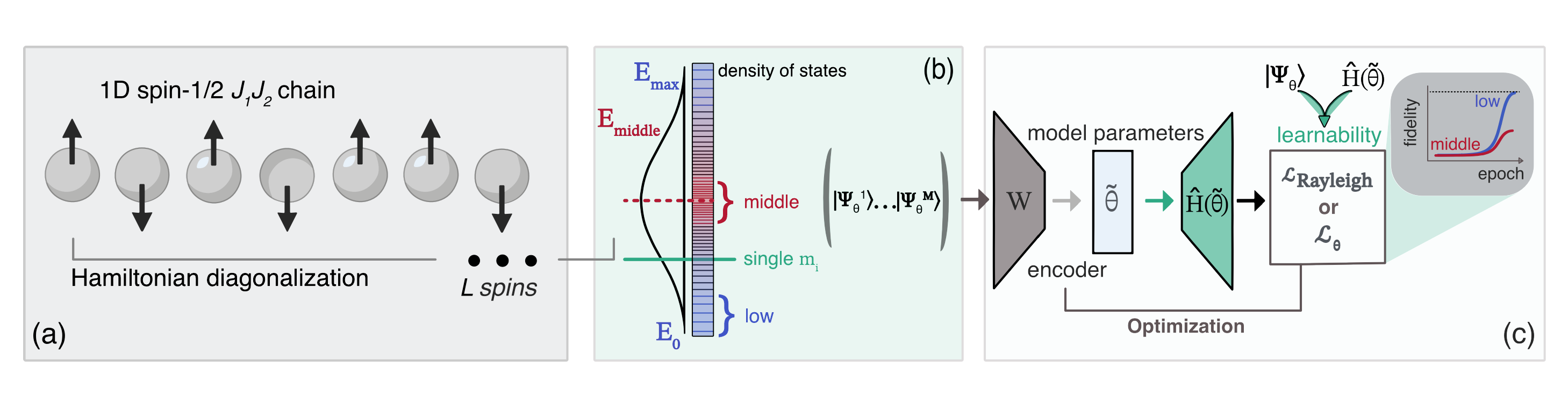}
    \caption{
        Schematic overview of our autoencoder setup for exploring learnability. (a) One-dimensional $J_1J_2$ spin-$1/2$ chain $\hat{H}(\theta)$ [Eq.~\eqref{eq:hamil}], specified by parameter vector $\theta$, is diagonalized to obtain a full set of eigenstates $\smash{\{\ket{\Psi_{\theta}^{m_i}}\}}$ and corresponding eigenvalues $\smash{\{E_\theta ^{m_i}\}}$. (b) Subsets of eigenstates are selected in one of three ways: (i) {\color{blue} low}-energy sector, (ii) {\color{red} middle} of the spectrum, (iii) {\color{darkgreen} single} eigenstate $m_i$. These states constitute the input to the network. (c) The encoder network uses quantum states to infer parameters $\smash{\tilde{\theta}}$; this latent representation is used by the decoder to obtain a reconstruction of the Hamiltonian.  We train using the physically-motivated loss function, $\smash{\mathcal{L}_{\rm Rayleigh}}$ [Eq.~\eqref{eq:rayleigh_loss}], and use the metric $\smash{\mathcal{L}_{\theta}}$ [Eq.~\eqref{eq:theta_loss}] to evaluate the final reconstruction. The expected behavior for the fidelity (which quantifies the overlap between input and reconstructed states) is shown in the cartoon:  As the losses decrease during training, the fidelity should increase. This behavior is expected to vary across the spectrum, with higher fidelities in the low-energy sector than in the middle of the spectrum.
    }
    \label{fig:sketch_idea}
\end{figure*}

In this work, we propose \emph{learnability} as a framework for characterizing the information encoded in many-body eigenstates.  
Specifically, it measures how accurately a fixed ML model can reconstruct the underlying Hamiltonian parameters from a subset of eigenstates.  While the absolute numerical precision of such reconstructions depends on architectural choices and training protocols, we are primarily interested in relative trends, such as the variation of learnability across the spectrum, which reflect intrinsic physical properties of the system. 

We explore the learnability framework using a Heisenberg ($\text{XXZ}$) spin chain including next-nearest-neighbor interactions.  
We examine how well Hamiltonians can be inferred directly from eigenstates chosen from different parts of the many-body spectrum.  The overall setup and idea are illustrated schematically in Figure \ref{fig:sketch_idea}.  This offers an ML-based approach for measuring how information is distributed across the spectrum, and the extent to which eigenstates in different parts of the spectrum preserve information about the underlying interactions.

Our results demonstrate that, for local Hamiltonians, a small number of low-lying eigenstates suffices to accurately reconstruct the model parameters. This information rapidly deteriorates as one moves toward the center of the spectrum, where eigenstates are difficult to distinguishable from random vectors even at modest system sizes. These findings establish a clear connection between the spectral position and the information-theoretic limits of the reconstruction.

In Sec.~\ref{sec:model}, we first introduce a generic model family under consideration and their latent representation $\theta$. We switch to learnability in Sec.~\ref{sec:learnability}, defining the protocols [Sec.~\ref{sec:intro:protocol}], the general network architecture in Secs.~\ref{sec:intro:encoder} and~\ref{sec:intro:decoder}, and the physically motivated loss function in Sec.~\ref{sec:loss}. Finally, we present an analysis of the results in Sec.~\ref{sec:results} and summarize our findings and related scenarios in Sec.~\ref{sec:conclusions}. Additional details and extended insights into the results are provided in Appendices~\ref{app:details}~and~\ref{app:extended}.
\section{Model Hamiltonians}\label{sec:model}

We focus on one-dimensional systems with local interactions.  Specifically, we consider a general family of spin-$1/2$ chain Hamiltonians,
\begin{eqnarray}\label{eq:hamil}
    & \hat{H}_{J_1, J_2} = & \sum _{i=1}^{L}J_1 \hat{\mathbf{h}}_{i,i+1} + \sum _{i=1}^{L}J_2 \hat {\mathbf{h}}_{i,i+2} + \hat{H}_{\rm loc} \;, \nonumber \\
    & \hat{H}_{loc} = & \sum _{i=1}^{L} \left [h_z^{(i)} \hat \sigma ^z _i + g_x^{(i)} \hat \sigma ^x _i\right ] \;, \nonumber \\ 
    & \hat{\mathbf{h}}_{ij} =& \hat \sigma ^x_i \hat \sigma ^x_j + \hat \sigma ^y_i\hat \sigma ^y_j + \Delta \hat \sigma ^z_i\hat \sigma ^z_j \;,
\end{eqnarray}
with nearest-neighbor $\smash{J_1}$ and next-nearest-neighbor $\smash{J_2}$ couplings subject to periodic boundary conditions (PBC), such that $\hat \sigma _{L+1}^{\tau} = \hat \sigma _1^\tau$ and $\hat \sigma _{L+2}^{\tau} = \hat \sigma _2^\tau$ for $\tau=x,y,z$. Here, $\hat \sigma _{i}^{\tau}$ denote Pauli operators acting on the local Hilbert space $\mathcal{H}_i$ with dimension $d_i=\rm dim(\mathcal{H}_i)=2$. The corresponding many-body Hilbert space $\mathcal{H}$ is given by a tensor product $\smash{\mathcal{H}=\otimes _{i=1}^L\mathcal{H}_i}$ with $\smash{\mathcal{D}=\mathrm{dim}(\mathcal{H})=\prod_i d_i=2^L}$. In addition to hopping terms $\smash{\hat{\mathbf{h}}_{ij}}$, the Hamiltonian $\hat{H}_{J_1, J_2}$ includes longitudinal magnetic fields $\smash{h_z^{(i)}}$ and transverse magnetic fields $\smash{g_x^{(i)}}$ through the local contribution $\hat H_{\rm loc}$.

We focus on a reduced parameter space and consider $J_1$ and $J_2$ as the only free parameters. All other couplings are fixed to $\smash{h_z^{(i)} = h_z = 0.5}$, $\smash{g_x^{(i)} = g_x = -0.2}$, and $\Delta = 1.0$, so the Hamiltonian is completely determined by a single vector
\begin{equation}\label{eq:theta}
    \theta = (J_1,J_2)\;,
\end{equation}
treated as an implicit (latent) parameter; see Fig.~\ref{fig:sketch_idea}(c). We focus on $L = 6$; results for other system sizes are presented in Appendix~\ref{app:details}. In the main text, we vary only $J_1$ while fixing $\smash{J_2 = 0.5}$, so that $\theta$ contains a single parameter.  In Appendix~\ref{app:extended}, we relax this constraint and allow $J_2$ to vary, so that $\theta$ is a vector in $\mathbb{R}^2$. 

To explicitly break residual symmetries and avoid spectral degeneracies, we introduce on-site perturbations,
\begin{equation}\label{eq:alteration}
    \begin{aligned}
        & h_z^{(1)} \to -h_z^{(1)},  \quad g_x^{(1)} \to -g_x^{(1)}, \\
        & h_z^{(\lfloor L/2 \rfloor)} \to h_z^{(\lfloor L/2 \rfloor)} - 0.1, \\
        & g_x^{(\lfloor L/2 \rfloor)} \to g_x^{(\lfloor L/2 \rfloor)} + 0.1  \;,
    \end{aligned}
\end{equation}
which break spatial and parity symmetries. By lifting these symmetries, we ensure that the eigenstates carry unique signatures of the global coupling $J_1$.

In its most general form, one could choose position-dependent couplings, so that $\hat{H}_{J_1,J_2}$ would be characterized by a parameter vector $\theta \in \mathbb{R}^{4L}$. Fully inferring all couplings is computationally demanding, requiring large datasets and highly expressive networks. Such expressivity diverts attention from our central question of learnability and could be the subject of future studies.

\section{Learnability Protocol\label{sec:learnability}}

To what extent do Hamiltonian eigenstates encode model information, and how does this depend on spectral position? In particular, do eigenstates uniquely determine the parameters of a given Hamiltonian family $\smash{\{ \hat{H}_{{\theta}} \}}$ [Eq.~\eqref{eq:hamil}], i.e., can one \emph{learn} the Hamiltonian from a few eigenstates?  How does this differ between low- and mid-spectrum states, when the latter often resemble random vectors? We now formalize the learnability as a  measure of how accurately Hamiltonian parameters can be inferred from eigenstates.

To this end, we consider a learning architecture, e.g., a neural network, that maps the eigenstates of a Hamiltonian $\smash{\{ \hat{H}_{{\theta}} \}}$ directly to its underlying parameters $\smash{\tilde \theta}$. Here, the notation $\tilde{\theta}$ denotes the predicted parameters, in contrast to the original corresponding parameters $\theta$.  For our calculations, we will use an autoencoder architecture~\cite{bank_autoencoders_2020}; however the concept is not specific to a particular architecture.  

Our aim in this paper is not practical reconstruction, constrained by the curse of dimensionality~\cite{gorban_highdimensionalbrain_2020}, but rather exploration of the limits of reconstruction and the dependence of these limits on the location in the spectrum. Accordingly, the specifics of the network used (outlined below) are not critical, but it is important to use the same network architecture in the comparison between different eigenstates.  

Specifically, we study a family of local Hamiltonians ${\hat{H}_\theta}$ parametrized by $\theta \in \mathbb{R}^\Theta$, and assess how well $\theta$ can be inferred from a subset of eigenstates following a chosen protocol, cf. Sec.~\ref{sec:spectral_position}. As explained in Section \ref{sec:model}, the parameter space has dimension $\Theta=1$ in the main text, and the $\Theta=2$ setup is treated in Appendix \ref{app:extended}. We do not assume access to parameter labels and rely solely on a physics-motivated loss function, described below in  Sec.~\ref{sec:loss}.


\subsection{Network architecture\label{sec:architecture}}

The network is based on an autoencoder architecture composed of an \emph{encoder} and a \emph{decoder}. Conceptually, the training workflow is organized into three main stages depicted in Fig.~\ref{fig:sketch_idea}. To obtain a reliable mapping and to analyze the statistical convergence of the learning process, we vary the number of training samples (realizations) across the parameter space, choosing $N_{\rm sam}=10^3,10^4,2\times10^4$, i.e., selecting parameters $\theta$ as described in Sec.~\ref{sec:model}. In general, a larger training set improves the model's ability to learn~\cite{Watkin_Rau_Biehl_RMP93, Halevy_Norvig_Pereira_IEEEintelsystems2009, pugliese_machinelearningbased_2021}.  Unless otherwise specified, we use $N_{\rm epo}=4000$ number of epochs for each training scenario.

\subsubsection{The encoder} \label{sec:intro:encoder}

The procedure is schematically outlined in Fig.~\ref{fig:sketch_idea}.  We start with exact diagonalization (ED) of our spin chain, Eq.~\eqref{eq:hamil}, for preselected parameters $\theta$.  The choice of parameters is discussed in Sec.~\ref{sec:results}.  The encoder takes as input a matrix of many-body eigenstates, $\Psi_{\theta}\in \mathbb{R}^{\mathcal{D} \times M}$, where $M$ denotes the number of eigenstates included in a single realization during training, and each eigenvector is a $D$-dimensional vector. In other words, the columns of $\smash{\Psi_{\theta}}$ are given by the eigenstates $\smash{\{\ket{\Psi_{\theta}^{m_i}}:\hat{H}_{\theta} \ket{\Psi_{\theta}^{m_i}} = E^{m_i}_\theta \ket{\Psi_{\theta}^{m_i}}\}}$, with $m_i = 1,\dots,M$.  The $M$ eigenstates are chosen from the many-body spectrum according to one of several possible protocols, shown schematically in   Fig.~\ref{fig:sketch_idea}(b) and described below in \ref{sec:intro:protocol}. 

The architecture of the encoder is intentionally kept minimal, in order to focus on the learnability and its spectral dependence. It consists of a multi-layer perceptron (MLP)~\cite{hornik_multilayer_1989}, which transforms the $M\mathcal{D}$ input features into a hidden representation parametrized with $w_H$ neurons (details in Appendix~\ref{app:details}). These are then mapped to the latent space $\smash{\tilde{\theta}}$, of dimension $\Theta$, as shown schematically in  Fig.~\ref{fig:sketch_idea}(c). The action of the encoder is described as a non-linear mapping $\mathcal{F}[\mathcal{W}]:\ \mathbb{R}^{\mathcal{D} \times M} \rightarrow \mathbb{R}^\Theta$ parameterized by weights $\mathcal{W}$, such that:
\begin{equation} \label{eq:encoder}
    \tilde{\theta}
    =\mathcal{F}[\mathcal{W}](\Psi_{ \theta})\;.
\end{equation}
For completeness, we describe the network details in Appendix~\ref{app:details}.

\subsubsection{Eigenstate selection protocols} \label{sec:intro:protocol}

In order to investigate how the difference between eigenstates in different parts of the spectrum is manifested through learnability, we consider several ways of selecting the eigenstates that we feed into the autoencoder. 

One protocol is a single-eigenstate sweep, indicated as ``single $m_i$'' (green) in the schematic of Fig.~\ref{fig:sketch_idea}(b).   A single eigenstate $M=1$ is used for each run; this eigenstate is chosen from across the lower half of the spectrum, $\smash{m_i=1,\dots,\mathcal{D}/2}$.  This is perhaps the most obvious way of comparing eigenstates in different parts of the spectrum.  Results from this selection protocol are discussed in Sec.~\ref{sec:spectral_position}.

It is also interesting to examine the dependence of learnability on the number $M$ of eigenstates fed into the autoencoder.  For this we use, for the low-energy eigenstates, the first $M$ Hamiltonian eigenstates, $\smash{m_i\in \{1,\dots,M\}}$.  This protocal is marked as `low' (blue) in  Fig.~\ref{fig:sketch_idea}(b).  Also, for the mid-spectrum eigenstates, we use a set of $M$ consecutive eigenstates centered around the eigenstate with index $m_{\rm av}$, whose energy is closest to the mean energy, $\smash{E_{m_{\rm av}}^{\theta}\approx E_{\rm av}\equiv \Tr \hat{H}_{\theta}/\mathcal{D}}$.  Thus $\smash{m_i\in [m_{\rm av} - \lfloor M/2\rfloor, m_{\rm av}+\lfloor M/2 \rfloor]}$. 
This protocol is marked as `middle' (red) in  Fig.~\ref{fig:sketch_idea}(b).

These protocols enable a systematic analysis of how the learnability varies with spectral position and the number of input eigenstates $M$.

\subsubsection{The decoder} \label{sec:intro:decoder}

The decoder maps latent variables $\tilde{\theta}$ to a Hamiltonian $\smash{\hat{H}_{\tilde{\theta}}=\tilde{\mathcal{F}}(\tilde{\theta})}$, whose eigenstates $\smash{\Psi_{\tilde{\theta}}}$ are required to match the input states $\smash{\Psi_\theta}$. 
This is enforced through a reconstruction loss $\smash{\mathcal{L}}$. The specific loss functions are detailed in Sec.~\ref{sec:loss}. Notably, the decoder does not contain trainable internal parameters as it is not a neural network.

Rather than directly comparing the eigenvectors, we define the loss $\mathcal{L}$ in a way that enforces the consistency between $\hat{H}_{\tilde{\theta}}$ and the input eigenstates.
Hence, the final step constitutes the construction of a physically motivated loss function, such as $\mathcal{L}_{\rm Rayleigh}$, to determine the learnability across the spectrum, as shown in Fig.~\ref{fig:sketch_idea}(c).

\subsection{Physically motivated loss function} \label{sec:loss}

There is no obvious choice of loss. To probe learnability, it should be invariant under unphysical properties (e.g., global phases), properly normalized, and computationally tractable without the need for repetitive diagonalization, which is  prohibitive due to exponential growth of the Hilbert space size $\mathcal{D}$.

Several natural options exist. One can directly compare eigenstates via fidelity $F$; which requires diagonalizing of $\smash{\hat{H}_{\tilde{\theta}}}$ in every optimization step, to get $\Psi_{\tilde\theta}$. Alternatively, one may minimize the residuals $\smash{|(\hat{H}_{\theta}-E_{\tilde \theta} ^{m_i})\ket{\Psi^{m_i}_{\tilde \theta}}|}$, which in turn need access to the target $\smash{\hat{H}_{\theta}}$
or, equivalently, to parameters $\theta$.   Both approaches introduce additional computational or informational overhead.

Instead, we adopt a \emph{Rayleigh-type} loss based on the action of the reconstructed Hamiltonian $\smash{\hat{H}_{\tilde{\theta}}}$ on the known eigenbasis $\smash{\Psi_{\theta}}$. 
It is inspired by its resemblance to the Rayleigh quotient, commonly used in variational algorithms~\cite{yuan_theoryvariational_2019,cerezo_variationalquantum_2021}, which quantifies how closely $\smash{\ket{\Psi^{m_i}_{\theta}}}$ approximates an eigenstate of $\smash{\hat{H}_{\tilde{\theta}}}$. Defining
\begin{equation} \label{eq:h_red}
    \hat{H}_{\mathrm{res}}[\Psi_\theta](\tilde{\theta}) = \Psi^{\dagger}_\theta \hat{H}_{\tilde{\theta}}\Psi_\theta \;,
\end{equation}
perfect reconstruction implies that $\smash{\hat{H}_{\mathrm{res}}}$ is diagonal with elements that match the target eigenvalues.
Hence, deviations from diagonality quantify the mismatch. We therefore define,
\begin{eqnarray}
    \mathcal{L}_{\mathrm{Rayleigh}}\!\left\{ \hat{H}_{\tilde{\theta}},\Psi_{\theta}\right\} =
    \frac{1}{\mathcal{N}M(M-1)}\sum_{i\neq j} |(H_{\mathrm{res}})_{ij}|^2
    + \nonumber \\
    \gamma\,
    \frac{1}{\mathcal{N}M}\sum_i |(H_{\mathrm{res}})_{ii} - E_i|^2
         \;,
    \label{eq:rayleigh_loss}
\end{eqnarray}
with $\mathcal{N}=\sum_i E_i^2/M + \varepsilon;\ \varepsilon \ll 1$. This penalizes off-diagonal weight and spectral mismatch (weighted by $\gamma=0.1$), while remaining invariant under global rescaling $\hat{H} \to \alpha \hat{H};\ \alpha \in \mathbb{R}$.

This construction avoids explicit diagonalization, reduces the computational cost from $\mathcal{O}(\mathcal{D}^3)$ to $\mathcal{O}(M \mathcal{D})$, and does not require access to the target parameters $\theta$. It is, in this sense, a more weakly biased probe of learnability. Rather than enforcing individual residuals, $\smash{\mathcal{L}_{\mathrm{Rayleigh}}}$ captures a collective deviation of $\smash{\hat{H}_{\tilde{\theta}}}$ from diagonality in the reference eigenbasis.

This choice of the loss function follows the paradigm of physics-informed neural network training~\cite{raissi_physicsinformedneural_2019,karniadakis_physicsinformedmachine_2021}, in which enforces the physical consistency of predictions, rather than a purely supervised approach.  Through optimization, the autoencoder learns a low-dimensional manifold $\theta$ of physically meaningful Hamiltonians. Consequently, the latent representation $\tilde{\theta}$ encodes the emergent local couplings that recreate the observed eigenstates.

Finally, the total information flow during the learning process, \ie optimizing $\mathcal{W}$ to minimize the $\mathcal{L}_\mathrm{Rayleigh}$ loss, can be summarized as
\begin{equation}\label{eq:final_loss}
\begin{split}
    \min_\mathcal{W}\;
    &\mathcal{L}_\mathrm{Rayleigh} \!
    \left\{
    \hat{H}_{\tilde{\theta}}\;,
    \Psi_\theta
    \right\},\\
    &\hat{H}_{\tilde{\theta}}=\tilde{\mathcal{F}}\!\left(\tilde{\theta}\right)\;,
    \\
    &\tilde{\theta}=\mathcal{F}[\mathcal{W}]\left(\Psi_\theta\right)\;,
\end{split}
\end{equation}
where $\mathcal{F}$ denotes the parameter predictor network (encoder), and $\tilde{\mathcal{F}}$ represents the decoder, i.e., the function that takes the parameters as input and yields the Hamiltonian as output.

For validation, we also define a parameter-space loss,
\begin{equation}
    \mathcal{L}_{\theta}
    =
    \sum_{\ell}
    \left(
        \tilde{\theta}^{(\ell)}
        -
        \theta^{(\ell)}
    \right)^2
    \;,
    \label{eq:theta_loss}
\end{equation}
which directly compares the inferred and reference parameters. Here, it is not used during training, but serves as an independent diagnostic of reconstruction accuracy; a supervised variant is discussed in Appendix~\ref{app:extended}.

\begin{figure}[t]
    \centering
    \includegraphics[width=1.0\linewidth]{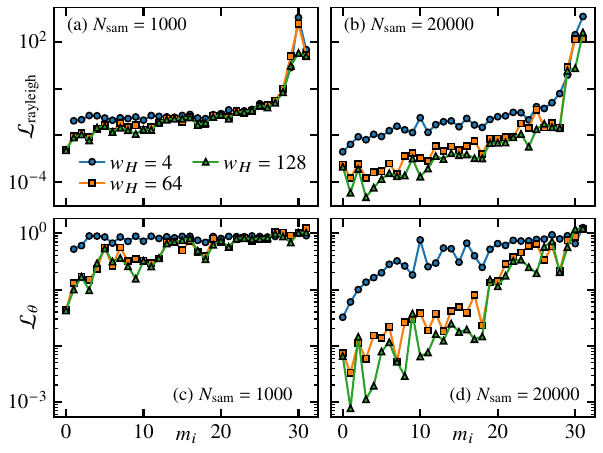}
    \caption{Dependence of the loss on the selected eigenstate index $m_i$ according to single state selection protocol. Columns correspond to different numbers of Hamiltonian realizations $N_{\mathrm{sam}}$. Panels (a)–(b) show the Rayleigh loss $\mathcal{L}_{\mathrm{Rayleigh}}$ [Eq.~\eqref{eq:rayleigh_loss}] used during training, while panels (d)–(f) display the reconstruction loss $\mathcal{L}_{\theta}$. Curves denote different hidden layer widths $w_H$.
    }
    \label{fig:sweep_spectrum}
\end{figure}

\section{Results\label{sec:results}}

We now quantify learnability in different parts of the spectrum, using the eigenstate selection protocols described in Sec.~\ref{sec:intro:protocol}.  We analyze how the selection of eigenstates, in particular their spectral position, affects the ability to reconstruct the Hamiltonian.

\begin{figure}[t]
    \centering
    \includegraphics[width=1.0\linewidth]{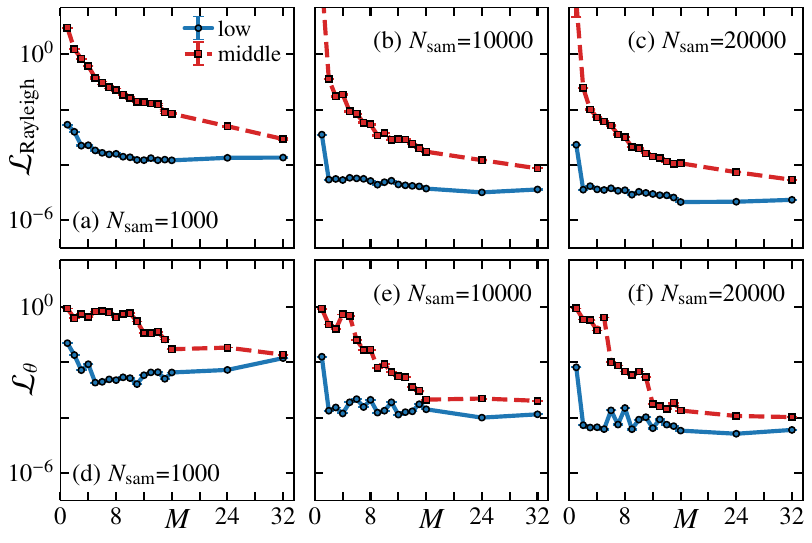}
    \caption{Systematic dependence of Hamiltonian reconstruction on the number of input eigenstates $M$ for two spectral protocols: (i) low-energy and (ii) middle of the spectrum. Columns correspond to different numbers of Hamiltonian realizations $N_{\mathrm{sam}}$. Panels (a)-(c) display the Rayleigh loss $\mathcal{L}_{\mathrm{Rayleigh}}$ [Eq.~\eqref{eq:rayleigh_loss}] used during training. Panels (d)-(f) show the parameter reconstruction loss $\mathcal{L}_{\theta}$ [Eq.~\eqref{eq:theta_loss}]. Assumed hidden layer width $w_H=128$.
    }
    \label{fig:sweep_M}
\end{figure}

\subsection{Reconstruction with a single state\label{sec:spectral_position}}

We begin by analyzing reconstruction from a single eigenstate ($M=1$) that is swept across the many-body spectrum, where we vary the position $m_i$ (schematic in Fig.~\ref{fig:sketch_idea}).  As illustrated in Fig.~\ref{fig:sweep_spectrum}, both the training and evaluation losses undergo a pronounced crossover. 
Eigenstates near the spectral edges allow for noticeably more accurate recovery of the Hamiltonian parameters $\theta$, and the reconstruction error increases steadily and substantially as one moves toward the center of the spectrum (larger $m_i$).  
The effect is seen equally well in the training loss $\mathcal{L}_{\mathrm{Rayleigh}}$ (top) and in the test/prediction error shown by the loss $\mathcal{L}_{\theta}$ (bottom).

The left and right panels of Fig.~\ref{fig:sweep_spectrum} correspond to different numbers of training samples $N_{\rm sam}$.  The change of learnability across the spectrum is visible in both cases.  The individual curves within each panel represent different network sizes, parametrized by the hidden layer size $w_H$.  For large enough $N_{\rm sam}$, we see an additional aspect of learnability --- at the spectral edge (low energy eigenstates), the reconstruction improves substantially with increasing $w_H$, while there is little or no improvement for the mid-spectrum eigenstates.

\subsection{Reconstruction with selected $M$ eigenstates \label{sec:multiple_states}}

\begin{figure}[t]
    \centering
    \includegraphics[width=1.0\linewidth]{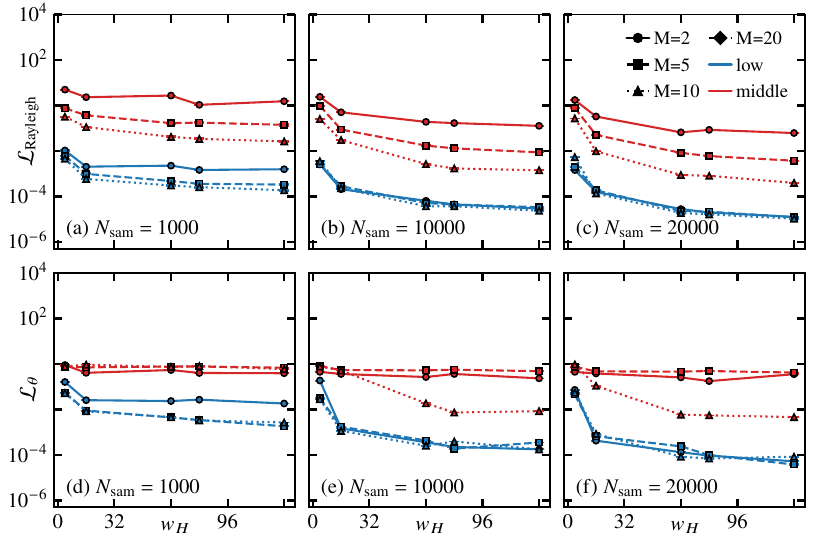}
\caption{
Dependence of the loss 
on the neural-network parametrization $w_H$ 
for two spectral protocols: (i) low-energy and (ii) middle of the spectrum. Columns correspond to different numbers of Hamiltonian realizations $N_{\mathrm{sam}}$. Panels (a)–(c) show the Rayleigh loss $\mathcal{L}_{\mathrm{Rayleigh}}$ [Eq.~\eqref{eq:rayleigh_loss}] used during training, while panels (d)–(f) display the reconstruction loss $\mathcal{L}_{\theta}$. Curves correspond to different numbers of input eigenstates $M$.
    }
    \label{fig:sweep_hidden}
\end{figure}

To isolate the role of spectral information and its improvement with increasing information, we now vary the number of input eigenstates $M$ and their spectral location, while keeping the number of hidden parameters fixed at $w_H=128$. We contrast low-energy states, which exhibit structured correlations and reduced entanglement, with mid-spectrum states, which are closer to random vectors.  The two eigenstate selection protocols are described in Sec.~\ref{sec:intro:protocol} and marked in Fig.~\ref{fig:sketch_idea} as `low' and `middle' respectively.

Fig.~\ref{fig:sweep_M}(a)-(c) tracks training performance with Rayleigh loss $\mathcal{L}_{\mathrm{Rayleigh}}$ [Eq.~\eqref{eq:rayleigh_loss}], while Fig.~\ref{fig:sweep_M}(d)-(f) reports post-training evaluation using $\mathcal{L}_{\theta}$ [Eq.~\eqref{eq:theta_loss}]. The columns correspond to different numbers of Hamiltonian realizations $N_{\rm sam}$. For the low-energy protocol, the metrics rapidly converge, with reconstruction errors systematically decreasing as $M$ increases. In stark contrast, training on mid-spectrum states fails to yield accurate reconstruction, regardless of $N_{\mathrm{sam}}$, unless $M$ approaches half of the spectrum $M\rightarrow \mathcal{D}/2$. This indicates a suppression of accessible local operator information in the highly entangled bulk.

Next, we investigate the impact of network capacity in Fig.~\ref{fig:sweep_hidden}. As the width of the hidden layer $w_H$ is varied, cf. Sec.~\ref{sec:architecture}, reconstruction yields progressively better results. However, while increasing the network size speeds up convergence for edge states, it does not remedy the lack of information in the middle of the spectrum. This further confirms that failure in the bulk is governed by intrinsic properties of the eigenstates, rather than limitations of model expressivity (see also Appendix~\ref{app:details}).

\begin{figure}[t]
    \centering
    \includegraphics[width=1.0\linewidth]{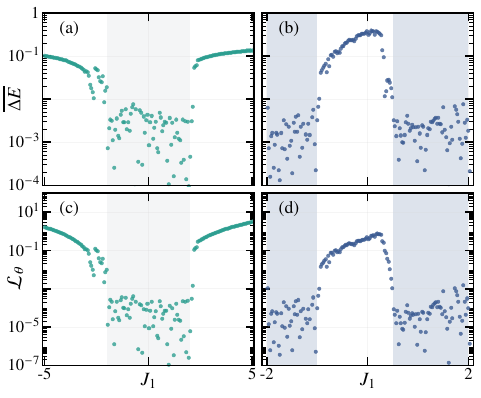}
    \caption{Generalization across domains for $M=5$ low-energy eigenstates at $L=6$. The network ($w_H=128$) is trained on two disjoint intervals of $\theta$, namely (a),(c) $J_1\in [-2, 2]$ and (b),(d) $J_1 \in [-2,-1]\cup[0.5,2]$. (a),(b) Relative eigenvalue error $\overline{\Delta E}$ [Eq.~\eqref{eq:eigval_err}]. (c),(d) Parameter evaluation loss $\mathcal{L}_{\theta}$ [Eq.~\eqref{eq:theta_loss}].
    }
    \label{fig:generalization_hole}
\end{figure}

We further investigate generalization in Fig.~\ref{fig:generalization_hole} by training the MLP ($w_H=128, M=5$) on two separate parameter intervals, such that either the middle interval [Fig.~\ref{fig:generalization_hole}(a),(c)] or the outer interval [Fig.~\ref{fig:generalization_hole}(b),(d)] is omitted from the training data. The model is then evaluated over the entire parameter range. In Fig.~\ref{fig:generalization_hole}(a),(b), we compare the predictions to the exact spectrum using the spectral error
\begin{equation} \label{eq:eigval_err}
    \overline{\Delta E} = \frac{1}{\mathcal{D}}\sum_{m_i=1}^{\mathcal{D}} \frac{|E_i - \tilde{E}_i|}{E_{\max} - E_{0}}\;,
\end{equation}
computed for each available $\theta$. Likewise, Fig.~\ref{fig:generalization_hole}(c),(d) show the parameter evaluation loss $\mathcal{L}_\theta$ [Eq.~\eqref{eq:theta_loss}] as a function of $J_1$.

Although the network recovers well the parameters within the training domain, both the energy discrepancy $\overline{\Delta E}$ [Eq.~\eqref{eq:eigval_err}] and the parameter error $L_\theta$ [Eq.~\eqref{eq:theta_loss}] increase markedly in the excluded region. This indicates that learnability does not automatically imply reliable interpolation and that, even when low-energy states contain sufficient information, simple architectures may still struggle to generalize over the entire parameter space.  Thus, generalization emerges as a distinct problem from learnability.
However, it should be noted that, in the region adjacent to the training domain, Fig.~\ref{fig:generalization_hole} shows a clearly lower prediction error that increases gradually, indicating some ability of the encoder to generalize beyond the training set.

\section{Discussion\label{sec:conclusions}}

\subsection{Context and related work}

ML has become a complementary approach to probe physical systems, for instance by identifying quantum correlations such as entanglement~\cite{deng_quantumentanglement_2017,Chen2022,koutny_deeplearning_2023,Pawlowski2024, feng_quantifyingquantum_2024,huang_directentanglement_2025} and discord~\cite{Krawczyk2024}. More recently, increasing attention has turned to questions of interpretability~\cite{zhang_interpretingmachine_2020,valenti_scalablehamiltonian_2022,wetzel_interpretablemachine_2025} and physics-informed ML~\cite{raissi_physicsinformedneural_2019,karniadakis_physicsinformedmachine_2021}. In several cases, ML has been used as a diagnostic, matching or surpassing conventional indicators~\cite{vannieuwenburg_learningphase_2017,zhang_machinelearning_2018,rodriguez_nieva_identifyingtopological_2019}. This suggests that ML can define data-driven indicators, providing a new class of probes on top of established theoretical approaches.

Our central question was how much information about a Hamiltonian is encoded in its eigenstates. From a practical perspective, this is addressed by Hamiltonian learning (HL)~\cite{dasilva_practicalcharacterization_2011,hentschel_machinelearning_2010,hentschel_efficientalgorithm_2011,
sergeevich_characterizationqubit_2011,bairey_learninglocal_2019,chertkov_computationalinverse_2018,gu_practicalhamiltonian_2024}. Its goal is to recover the governing operator, most often using measurements acquired either in a steady state~\cite{Feng2024,Lupi2025,Henderson2023,Khosravian_2024,liu2025,pawlowski2026} or during quantum dynamics~\cite{Bonizzoni2022,Karjalainen2023,Mirani2024,StilckFranca2024,Hangleiter2024}. HL has been successfully applied in several areas, e.g., in analyzing transport measurements~\cite{Koch2023,Taylor2024disorderlearning,Thamm2024,Taylor2025analysis,krawczyk2026ai,pawlowski2026}, inferring tight-binding Hamiltonians~\cite{Gu2024,Choudhary2025}, and characterizing spin models~\cite{Lupi2025,Bonizzoni2022} or quantum circuits~\cite{jerbi_parametrizedquantum_2021,acampora_deepneural_2021,skolik_quantumagents_2022}. A key challenge is the uniqueness of the Hamiltonian compatible with a given state. Most generic local Hamiltonians can be reconstructed from measurements of a single eigenstate~\cite{qi_determininglocal_2019}; see also Ref.~\cite{garrison_doessingle_2018a} for related arguments. Subsequent work extended this to efficient reconstruction from local observables~\cite{bairey_learninglocal_2019a} and quantified the number of independent constraints extractable from an eigenstate~\cite{zhou_learningsymmetric_2024}. Similar uniqueness guarantees have been obtained for Gibbs states and nonequilibrium steady states~\cite{bairey_learningdynamics_2020,haah_learningquantum_2024}, with the feasibility of reconstruction linked to locality and the decay of correlations~\cite{anshu_sampleefficientlearning_2021}. Nonetheless, this perspective does not fully resolve how the available information depends on the spectral characteristics of the states.

Several (already) conventional diagnostics characterize the structure of many-body eigenstates, such as entanglement entropy~\cite{hastings_arealaw_2007,eisert2010colloquium,bianchi_volumelawentanglement_2022a}, correlation functions~\cite{hahn_eigenstatecorrelations_2024,pappalardi_eigenstatethermalization_2022a,schweigler_experimentalcharacterization_2017,wang_eigenstatethermalization_2026}, \emph{compressibility}~\cite{Zala2022,Luchnikov2019}, and effective dimension measures~\cite{detomasi_multifractalitymeets_2020,roy_fockspaceanatomy_2021a,schreiber_fractalcharacter_1985}, depending on the community. In principle, eigenstates determine the Hamiltonian up to spectral ambiguities~\cite{vonneumann_mathematicalfoundations_2018,reed_methodsmodern_1972,dAlessio16}. All of these measures, however, consistently indicate a sharp contrast between low-energy and mid-spectrum states~\cite{dAlessio16,bianchi_volumelawentanglement_2022a}.

We have established a clear quantitative link between the spectral position of many-body eigenstates and the learnability of their parent Hamiltonian.  We demonstrated that low-energy eigenstates preserve enough structure to reliably reconstruct local couplings, whereas this information is gradually and intrinsically lost as one moves toward the middle of the spectrum. This loss of learnability is largely independent of network capacity and instead reflects fundamental properties of many-body eigenstates, in line with their crossover to thermal, pseudo-random behavior~\cite{dAlessio16}. Our findings show that the feasibility of extracting the Hamiltonian structure is not uniform throughout the spectrum but dictated by the underlying physics of the states themselves.

\subsection{Defining learnability (quantitatively)}

In this work, we have used the term \emph{learnability} semi-qualitatively, i.e., not as a single number assigned to each eigenstate but rather as an framework governing overall behavior.  The specific values of loss used to analyze the concept depend on the learning architecture and loss function.  We expect the reported results and trends, and the underlying intuitions, to be broadly independent of these choices.  

In addition to the conceptual understanding of the framework, it is also interesting to consider possible mathematical definitions.  We propose one below. 

The learning ability emerges if there is anything to be learned by a capable model. In this context, one can formulate a more precise notion of learnability in terms of how the reconstruction fidelity can change with increasing model capacity. Namely, one could define
\begin{equation}\label{eq:learnability}
    \Delta\mathcal{L}=\mathcal{L}(\mathcal{W}_0^*) -\min_{\mathcal{W}^* \in \mathcal{C}} \mathcal{L}(\mathcal{W^*})\;,
\end{equation}
where $\mathcal{C}$ denotes a class of models of increasing capacity, i.e., parameter count (here, the hidden layer width $w_H$), $\mathcal{W^*}$ means trained models and $\mathcal{W}_0^*$ represents a baseline reference with the lowest fidelity. 
A reconstruction task is \emph{learnable}, i.e., is able to learn from eigenstates, if $\mathcal{L}(\mathcal{W^*})$ decreases systematically as the capacity of the model increases, ultimately approaching zero. Conversely, if $\mathcal{L}(\mathcal{W}^*)$ remains close to the reference value $\mathcal{L}(\mathcal{W}_0^*)$, the data may not contain sufficient information to uniquely determine the target parameters, regardless of increases in model capacity. Equivalently, one can quantify learnability via the \emph{loss gap} clearly visible in Fig.~\ref{fig:sweep_spectrum}(d), defined as the difference between the attainable losses of low- and high-capacity models. Hence, a nonzero gap indicates learnability.

\subsection{Perspective}

This work establishes \emph{learnability} as a new information-theoretic framework for quantum many-body systems. It shifts the role of ML from being merely a computational tool to serving as a probe of the physical structure.  In this way, our method offers a complementary route, based on  learning architectures that have become available in recent years, to investigate how information is stored, preserved, and erased in complex quantum systems.  
%

The present work opens up a host of open questions and further research avenues:

(-) Future work may address the notion of learnability across a variety of many-body problems.  We have chosen to focus on a single chaotic system.  How does the spectral dependence of learnability look like for integrable systems, which are known to resist the standard thermalization paradigm \cite{dAlessio16}?  Similar questions arise for many-body-localized systems and systems with many-body scars, cases where mid-spectrum eigenstates are not necessarily random-like.  

(-) The numerical precision with which the Hamiltonian is reconstructed depends on the particular ML architecture.  Relative trends, such as the dependence on spectral position or the dependence on the number $M$ of eigenstates used, reflect physical properties of the system, and thus should be robust across different setups.  It would be worthwhile to explore this robustness, i.e., to find out if anything changes qualitatively for a different learning architecture.  A related question is whether particular architectures might be more suitable than others for characterizing learnability.  

(-) In addition to many-body quantum systems, it would be interesting to apply the learnability framework to single-particle quantum systems, e.g. hard-wall or soft-wall quantum billiards.  While ETH or thermalization might not be directly relevant to these systems, eigenstate properties still vary across the spectrum, and there are significant differences between eigenstates of chaotic, integrable and intermediate systems.  Some of this physics might be fruitfully probed using the learnability framework.

(-) More specifically, it would be of interest to examine how learnability connects to the entanglement structure of eigenstates, the degree to which their information can be compressed~\cite{Zala2022}, and the extent to which neural networks can generalize the underlying structure of eigenstates beyond the training domain.\\

\textit{Data availability.--}
Research data and snipped code associated with this article are available on Zenodo~\cite{zenodo}.

\acknowledgments 
MK acknowledges support from the National Science Centre (Poland) under Grant No. 2024/53/B/ST3/02756.  MH acknowledges support from the Deutsche Forschungsgemeinschaft under grant SFB 1143 (project-id 247310070).

\appendix

\section{Definition of entropies}\label{app:entropy}

In the main text, specifically in Fig.~\ref{fig:fig0}, we illustrated how the properties of eigenstates depend on their spectral position by presenting the behavior of the entanglement entropy [panels (c),(d)] and the participation entropy [panels (e),(f)] as functions of the rescaled energy. For completeness, in this section we provide the definitions of these quantities.

For pure quantum states $\ket{\Psi}$, taken here to be a Hamiltonian eigenstate, we consider the bipartite entanglement entropy of a subsystem $A$ consisting of $L_A=L/2$ contiguous spins in the chain. The reduced density matrix is obtained by tracing out the complementary subsystem $B$,
\begin{equation}\label{eq:rdm}
    \hat{\rho}_A = \Tr _B |\Psi \rangle \langle \Psi |\;.
\end{equation}
The von Neumann entanglement entropy of the subsystem $A$ is
\begin{equation}\label{eq:vnentropy}
    S_{\rm vN} = -\Tr \hat \rho _A \ln \hat \rho _A \;,
\end{equation}

We additionally describe the eigenstates using the generalized inverse participation ratio (IPR),
\begin{equation}\label{eq:ipr}
    P_q^{-1}(\ket{\Psi}) = \sum_{i=1}^{\mathcal{D}}|\braket{i}{\Psi}|^{2q}\;.
\end{equation}
where $\{\ket{i}\}$ denotes the computational (spin) basis. The corresponding participation entropy is defined as
\begin{equation}\label{eq:part:ent}
    S_{\mathrm{part}}^{(q)} = \frac{1}{1-q} \ln P_q^{-1}(\ket{\Psi})\;.
\end{equation}
In the limit $q \to 1$, which we consider, one obtains the Shannon participation entropy,
\begin{equation}\label{eq:shannon}
    S_{\mathrm{part}} = - \sum_{i=1}^{\mathcal{D}} |\braket{i}{\Psi}|^2 \ln |\braket{i}{\Psi}|^2\;.
\end{equation}
%

\section{Neural network details} \label{app:details}

\begin{figure}[!ht]
    \centering
    \includegraphics[width=1.0\linewidth]{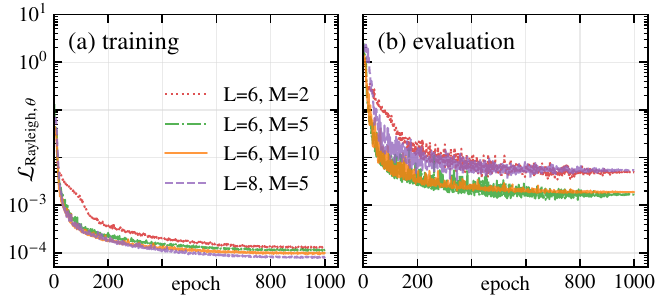}
    \caption{(a) Training history using $\mathcal{L}_{\rm{Rayleigh}}$ [Eq.~\eqref{eq:rayleigh_loss}], evaluated in terms of (b) $\mathcal{L}_\theta$ [Eq.~\eqref{eq:theta_loss}] for representative system sizes $L=6,8$ and numbers of input eigenstates $M=2,5,10$. In all cases, $w_H=128$ and $N_{\rm epo}=1000$ and \emph{low} protocol is used; cf. Sec.~\ref{sec:intro:protocol}.
    } 
    \label{fig:app:training}
\end{figure}

Here, we present a more thorough description of the network architecture used to obtain the main results of our study. To realize our methodology, depicted in Fig.~\ref{fig:sketch_idea}, we employ the PyTorch framework~\cite{paszke_pytorchimperative_2019} with single-precision floating-point arithmetic. All computations are carried out on an NVIDIA GeForce 3050 Ti graphics processing unit (GPU). Each optimization run is done by splitting the full dataset of size $N_{\rm sam}$ into a training set with $N_{\rm train}=0.7N_{\rm sam}$ samples and a validation set with $N_{\rm train}=0.3N_{\rm sam}$ samples. We use AdamW optimizer, though we do not anticipate our findings to be highly sensitive to this choice. The code is publicly available upon request.

The \emph{encoder} (the only trainable component of the network), cf. Fig.~\ref{fig:sketch_idea}(b), is implemented as a lightweight point-wise multi-layer perceptron (MLP) acting independently on the amplitudes of the input eigenstates $\Psi_\theta \in \mathbb{R}^{D \times M}$. Each vector is first mapped to a hidden representation of width $w_H$ via a linear layer, followed by normalization and a SiLU activation. This is further processed by a \emph{residual block} composed of two linear layers with normalization and a skip connection $x \mapsto \mathrm{SiLU}(x + f(x))$. The resulting features are finally projected to the latent parameters $\tilde{\theta}$ using a shallow readout network. In this setup, the model capacity is primarily controlled by $w_H$, enabling a systematic study of learnability as a function of architectural complexity in Sec.~\ref{sec:results}, without distracting from the central findings of this work.

\begin{figure}[t]
    \centering
    \includegraphics[width=1.0\linewidth]{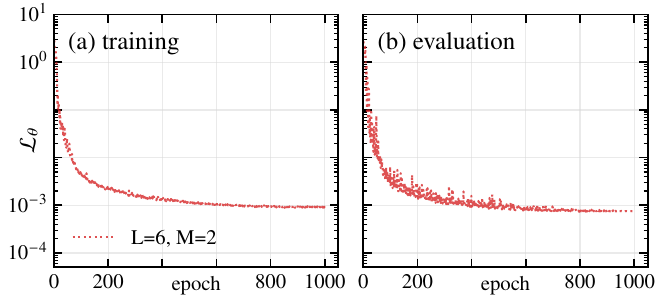}
    \caption{(a) Training history using the supervised loss $\mathcal{L}_\theta$ [Eq.~\eqref{eq:theta_loss}], with (b) corresponding evaluation on testing dataset, for system size $L=6$ and $M=2$ input eigenstates. Here, $w_H=128$, $N_{\rm epo}=1000$, and the \emph{low} protocol is used; cf. Sec.~\ref{sec:intro:protocol}.
    } 
    \label{fig:app:training_theta}
\end{figure}

For completeness, in Fig.~\ref{fig:app:training} we additionally present representative training curves that illustrate the optimization dynamics for selected training runs performed with the \emph{low} protocol; see Sec.~\ref{sec:intro:protocol} of the main text. Fig.~\ref{fig:app:training}(a) displays the $\mathcal{L}_{\rm Rayleigh}$ curve [Eq.~\eqref{eq:rayleigh_loss}] used during optimization, while Fig.~\ref{fig:app:training}(b) shows the corresponding $\mathcal{L}_\theta$ values [Eq.~\ref{eq:theta_loss}] evaluated on the validation set. Both curves exhibit stable convergence across different configurations, with faster decay observed for smaller values of $M$ and lower system size.

Finally, to verify that our results are not specific to the choice of the Rayleigh loss $\mathcal{L}_{\rm Rayleigh}$ [Eq.~\eqref{eq:rayleigh_loss}], we perform a fully supervised training, this time using the parameter loss $\mathcal{L}_\theta$ [Eq.~\eqref{eq:theta_loss}]. As shown in Fig.~\ref{fig:app:training_theta}, this yields qualitatively identical behavior. 

\section{Extended training regime} \label{app:extended}

In the main text, we restricted the Hamiltonian family, see Eq.~\eqref{eq:hamil}, to a single free parameter $J_1$ in order to isolate the spectral dependence of learnability. Here, for completeness, we extend the analysis to a two-parameter setting, allowing both nearest- and next-nearest-neighbor couplings $(J_1, J_2)$ to vary, while keeping the remaining parameters fixed as in Sec.~\ref{sec:model}.

\begin{figure}[!b]
    \centering
    \includegraphics[width=1.0\linewidth]{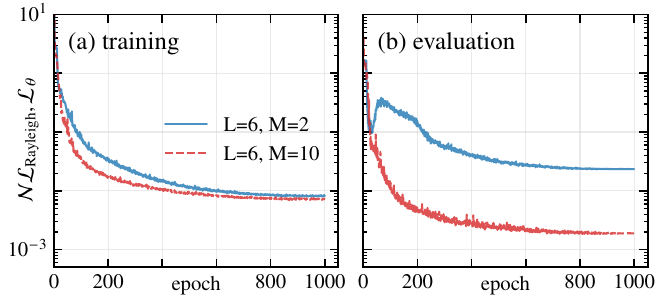}
    \caption{(a) Training history using $\mathcal{N}\mathcal{L}_{\rm{Rayleigh}}$ [Eq.~\eqref{eq:rayleigh_loss}], evaluated in terms of (b) $\mathcal{L}_\theta$ [Eq.~\eqref{eq:theta_loss}] for system size $L=6$ and numbers of input eigenstates $M=2,10$. In all cases, $w_H=128$ and $N_{\rm epo}=1000$ and \emph{low} protocol is used; cf. Sec.~\ref{sec:intro:protocol}. The model simultaneously infers both couplings $(J_1,J_2)$.
    } 
    \label{fig:app:training_j1j2}
\end{figure}
The training protocols and network architecture remain unchanged. The encoder now maps the input eigenstates $\smash{\Psi_\theta}$ into a two-dimensional latent space $\smash{\tilde{\theta} = (\tilde{J}_1, \tilde{J}_2)}$. Here, the inference task requires disentangling several competing interaction terms from the same set of eigenstates. We restrict the analysis to the low-energy protocol.

As illustrated in Fig.~\ref{fig:app:training_j1j2}, the optimization exhibits stable convergence, displaying behavior that is qualitatively similar to the single-parameter scenario.  This indicates that the observed limitations of learnability persist beyond the minimal setup and are not merely a consequence of the reduced parameter space. It should be noted, however, that the training process clearly becomes more difficult, as reflected by a higher final evaluation loss $\mathcal{L}_\theta$ in Fig.~\ref{fig:app:training_j1j2}(b).

\bibliographystyle{apsrev4-2}
\bibliography{references}

\end{document}